\documentclass[apsprd]{revtex4-1}
\usepackage{amsfonts}
\usepackage{amsmath}
\usepackage{amssymb}
\usepackage{graphicx}
\usepackage{color}

\def\be{\begin{equation}}
\def\te{\end{equation}}
\def\ee{\end{equation}}
\def\ba{\begin{eqnarray}}
\def\bea{\begin{eqnarray}}
\def\nn{\nonumber\\}
\def\tea{\end{eqnarray}}
\def\ea{\end{eqnarray}}
\def\eea{\end{eqnarray}}

\begin{document}

\title{Linearized dispersion relations in viscous relativistic hydrodynamics}

\author{Guillermo Perna}
\email{guillermoezequielperna@gmail.com}
\author{Esteban Calzetta}
\email{calzetta@df.uba.ar}
\affiliation{Departamento de F\'isica, Facultad de Ciencias Exactas y Naturales, Universidad de Buenos Aires and IFIBA, 
CONICET, Ciudad Universitaria, Buenos Aires 1428, Argentina}

\pacs{52.27.Ny, 52.35.-g, 47.75.+f, 25.75.-q}

\begin{abstract}
We compute the dispersion relations for scalar, vector and tensor modes of a viscous relativistic fluid, linearized around an equilibrium solution, for a divergence type theory (which, in the linearized theory, includes Israel-Stewart and anisotropic hydrodynamics as particular cases) and contrast them to the corresponding results derived from kinetic theory under the relaxation time approximation, and from causal first order theories. We conclude that all approaches give similar dynamics for the scalar and vector modes, while the particular divergence type theory presented here also contains propagating damped tensor waves, in agreement with kinetic theory. Non hydrodynamic tensor modes are also a feature of holographic fluids. These results support the application of hydrodynamics in problems involving the interaction between fluids and gravitational waves.
\end{abstract}
\maketitle

\section{Introduction}

Recent developments in relativistic heavy ion collisions \cite{Rom-Rom-19} and cosmology \cite{donough-20} have brought attention to the physics of relativistic viscous fluids \cite{CalHu08,RZ13}, particularly since the realization that hydrodynamical models act as an attractor to more complex physics even on short times scales \cite{attractor1,attractor2,attractor3,Rom18,Strickland18b,Chattopadhyay20,Kurkela20}. However, progress has been impaired by the fact that, unless the situation for non relativistic fluids described by the Navier-Stokes equations, no single approach to relativistic viscous fluids has achieved consensus status in the community. This is not a matter of ``right'' vs. ``wrong'' but rather that different approaches best capture some aspects of the complex physics of relativistic fluids.

Given this situation, it is important to develop tests where the predictions of different approaches may be contrasted, thereby helping to select the most adequate choice for a given physical problem. One strategy that has been extensively used in the literature is to apply different approaches to a problem which could also be solved using a more fundamental theory. In this sense, the Bjorken and Gubser models of the expanding fireball in a relativistic heavy ion collision have been a preferred choice \cite{BG1,BG2,Nor1,CanCal20}. In this case the more fundamental theory is kinetic theory under the relaxation time approximation \cite{AnWit74a,AnWit74b,BGK,TI10}, or else holographic fluids in an asymptotic region \cite{BRSSS}.

Another approach is to consider linearized perturbations of an equilibrium state, and to identify the propagating modes and their dispersion relations. Dispersion relations are known from kinetic theory under the relaxation time approximation \cite{Rom-Rom-19,Roma666,KurWie,Heller18,Heller21}. They may be found also from quantum field theories, in the weakly coupled limit trough a perturbative expansion in the coupling constant \cite{Moore}, or else in the infinite coupling limit for holographic fluids \cite{Rom-Rom-19,BRSSS,PolSonStar,SonStarBlackMagic,PolSonStar2,PolSonStar3,HerSon,KotSonStar,KotStar,ChesYaf,Withers,Grozdanov}. All these approaches give similar though not identical results. Our goal is to contrast these ``first principles'' dispersion relations to the ones obtained from hydrodynamics.

We shall consider only conformal theories with no conserved charges. Then the focus of interest of a hydrodynamic model is the energy-momentum tensor (EMT) $T^{\mu\nu}$, which satisfies the conservation law
\be
T^{\mu\nu}_{;\nu}=0
\label{cons}
\te
There is also an entropy flux $S^{\mu}$ which satisfies the ``Second Law''
\be
S^{\mu}_{;\mu}=\Sigma
\label{secondlaw}
\te
where $\Sigma\ge 0$ is the entropy production.

For an ideal fluid
\be
T^{\mu\nu}_{ideal}=\rho u^{\mu}u^{\nu}+p\Delta^{\mu\nu}
\label{ideal}
\te
where $\rho$ is the energy density, $u^{\mu}$ is the velocity, restricted to the shell $u^2=-1$ (we work with signature $(-,+,+,+)$ and natural units $\hbar = k_B = c = 1$), $\Delta^{\mu\nu}=g^{\mu\nu}+u^{\mu}u^{\nu}$ and $p$ is the pressure. For a conformal fluid $T^{\mu\nu}$ must be traceless, $T^{\mu}_{\mu}=0$, and so $p=\rho/3$. Since there are no conserved charges and therefore no chemical potentials we may define the entropy density $s$ and the temperature $T$ through

\be
\frac{\partial p}{\partial T}=s=\frac{p+\rho}T
\te
Then $\rho=\sigma T^4$, where $\sigma$ is a constant, $s=4\sigma T^3/3$, $S^{\mu}=su^{\mu}$ and $\Sigma=0$. 

In equilibrium the EMT of any fluid takes the ideal form Eq. (\ref{ideal}) \cite{I88}. Thereby in a weakly nonequilibrium state it is natural to write
\be
T^{\mu\nu}=T_0^{\mu\nu}+\Pi^{\mu\nu}
\label{viscous}
\te
where $T_0^{\mu\nu}$ has the ideal form eq. (\ref{ideal}) and    $\Pi^{\mu\nu}$ describes the viscous effects. For a conformal fluid $\Pi^{\mu\nu}$ must be traceless.

The problem is that now the four equations (\ref{cons}) are not enough to describe the evolution of the ten components of the symmetric tensor $T^{\mu\nu}$. This leads to two large families of theories. In the so-called ``first order'' theories (FOTs), constitutive relations are provided which restrict $\Pi^{\mu\nu}$ to be a given function of $T$, $u^{\mu}$ and their derivatives, thus keeping the number of degrees of freedom down to four. The so-called ``second order'' theories (SOTs), on the other hand, regard $\Pi^{\mu\nu}$, or a set of ``non equilibrium tensors'' from which $\Pi^{\mu\nu}$ may be computed, as dynamical variables on their own right, and provide supplementary equations of motion. 

Historically the first approaches to relativistic viscous fluids have been the Eckart and Landau-Lifshitz ones, which belong to the FOT class \cite{Eck40,LL59}. They provide covariant generalizations of the Navier-Stokes equations. They differ from each other in the definition of the fluid velocity $u^{\mu}$, which in the Eckart approach is defined from the flux of a conserved charge, while in the Landau-Lifshitz approach is defined form the energy flux, namely 

\be
T^{\mu\nu}u_{\nu}=-\rho u^{\mu}
\label{LL}
\te
which also defines $\rho$. Since the velocity apears in the constitutive relations for the viscous tensor $\Pi^{\mu\nu}$, the difference between the Eckart and Landau-Lifshitz theories is not just a matter of a choice of ``frame'', they actually are different theories \cite{Monnai19}. Since in this paper we shall only consider conformally invariant theories with no conserved charges, it is natural to restrict ourselves to the Landau-Lifshitz approach, to be discussed in more detail below

These first generation FOTs were proven to violate causality and to have no stable solutions \cite{HisLind1,HisLind2,HisLind4,HisLind5,HisLind6,VdL,PreRuRe20,PreRe20,Gavassino20}. However, it has been claimed that first order theories may be causal and stable if more general constitutive relations are considered \cite{Nor1,Koide07,Denicol09,Van09,VanBiro12,Nor2,Nor3,Kov19,HouKov20,Freistuhler21}. This claim has gained considerable attention, since first order theories are generally simpler that second order ones, and in particular easier to implement numerically \cite{Pandya21}. 

{ Concerning more general approaches such as BRSSS  \cite{Rom-Rom-19,BRSSS,BHMR08} and third order hydrodynamics \cite{J13,thirdorder}, which consists on writing the most general form for the energy-momentum tensor containing terms with up to a certain number of derivatives restricted by conformal invariance, the issue is whether the viscous energy momentum tensor is restricted to be proportional to the shear tensor, or else regarded as a hydrodynamic variable on its own. In the first case we obtain a theory within the FOT class, while in the second it becomes a SOT.}

The problem with the Eckart and Landau-Lifshitz approaches may be tracked down to the fact that, when defining the entropy production, some second order terms were retained while others were arbitrarily rejected \cite{I88}. This problem may be solved (or at least alleviated) by enlarging the set of degrees of freedom of the theory, and likewise introducing new terms in the entropy production. This leads to SOT approaches, such as Israel-Stewart \cite{Isr76,IS76,IS79a,IS79b,IS80,OH90}, Extended Thermodynamics \cite{PJC80,JRC80,JCL10}, DNMR \cite{DeKoRi10,BDKMNR11,DNNR11,DMNR12,DNMR12,DNBMXRG14,DN12,MNDR14,ND14}, Anisotropic Hydrodynamics \cite{Strick14a,Strick14b,FMRT15,FRST16,NMR17,BHS14,MNR16a,MNR16b}, and Divergence Type Theories (DTTs) \cite{LMR86,GL90,GL91,ReNa95,LeReRu18,Cal98,PRCal09,PRCal10,PRCal13,MGKanCal20,Cal21}. We shall focus on a particular implementation of the DTT paradigm, which, at the linearized level, contains the others as particular cases.

To the best of our knowledge, both FOTs and SOTs have been tested in Bjorken and Gubser backgrounds, where they successfully reproduce the results from kinetic theory under the relaxation time approximation \cite{Nor1,CanCal20,DFNR20}.

In this paper we will consider a conformal fluid in an equilibrium state in Minkowski space time, and compute the response of the EMT to a perturbation in the metric, assuming the dynamics is described by a DTT to be presented below. We shall compare the result with the same quantity as derived from a FOT, and from kinetic theory under the relaxation time approximation. We shall comment briefly on the corresponding result for quantum fields \cite{Moore,Rom-Rom-19,BRSSS,PolSonStar,SonStarBlackMagic,PolSonStar2,PolSonStar3,HerSon,KotSonStar,KotStar,ChesYaf,Withers,Grozdanov}.  In other words, we shall compute the propagator
\be
G^{\mu\nu\rho\sigma}\left[x,x'\right]=\left.\frac{\delta T^{\mu\nu}\left[x\right]}{\delta g_{\rho\sigma}\left[x'\right]}\right|_{g_{\mu\nu}=\eta_{\mu\nu}},
\label{props}
\te
The poles of the propagator as we approach the infrared limit indicate the propagating modes in the hydrodynamic limit and their dispersion relations. 

{The rest of the paper is organized as follows. In the next section we summarize the well known dispersion relations from kinetic theory \cite{Roma666,KurWie,Heller18,Heller21} (section (\ref{Kino})). Then in the following section we compute the EMT response in a divergence type theory (section (\ref{DTT})). We present some brief conclusiones in the final section.}

{For completitude we also present the relevant dispersion relations for ideal and Landau-Lifshitz fluids and causal FOTs \cite{Nor1} in an Appendix (section (\ref{CFOTs})).}

\section{Dispersion relations from kinetic theory}\label{Kino}

In this section we shall derive the dispersion relations from kinetic theory. We observe that the kinetic theory EMT propagators are computed in closed form in refs. \cite{Roma666,KurWie,Heller18,Heller21}. They display a complex analytic structure dominated by branch cuts. However, in the asymptotic regime where hydrodynamics is expected to hold, this analytic structure may be mimmicked by a suitable distribution of poles. Our interest lies in finding these equivalent poles in the $k^2\to 0$ limit, both the hydrodynamic ones and the longest living non hydrodynamic modes.

We consider an equilibrium state in Minkowski space time, whereby the metric $g_{\mu\nu}=\eta_{\mu\nu}=\mathrm{diag}\;\left(-1,1,1,1\right)$, the velocity $U^{\mu}$ and the temperature $T_0$ are constant. The EMT $T_0^{\mu\nu}$ takes the ideal form eq. (\ref{ideal}).  Without loss of generality we may assume $U^{\mu}=\left(1,0,0,0\right)$. We consider a fluctuation in the metric $\eta_{\mu\nu}\to g_{\mu\nu}=\eta_{\mu\nu}+h_{\mu\nu}$, and the corresponding change in the EMT $T^{\mu\nu}=T_0^{\mu\nu}+\delta T^{\mu\nu}$. Linearizing with respect to $h_{\mu\nu}$ we may read the propagator eq. (\ref{props}) from the relationship
\be
\delta T^{\mu\nu}\left[x\right]=\int\;d^4x'\;G^{\mu\nu\rho\sigma}\left[x,x'\right]h_{\rho\sigma}\left[x'\right]
\te
Four of this relationships are trivial, since they correspond to coordinate changes. If $x^{\mu}\to x'^{\mu}=x^{\mu}+\xi^{\mu}$, then $h_{\mu\nu}=-\xi_{\mu,\nu}-\xi_{\nu,\mu}$, and $\delta T^{\mu\nu}=\xi^{\mu}_{,\rho}T_0^{\rho\nu}+\xi^{\nu}_{,\rho}T_0^{\rho\mu}$. So, using $\Delta_0^{\mu\nu}=\eta^{\mu\nu}+U^{\mu}U^{\nu}$ we get
\be
\rho_0\left[\dot\xi^{\mu}U^{\nu}+\dot\xi^{\nu}U^{\mu}+\frac13\left(\Delta_0^{\mu\rho}\xi^{\nu}_{,\rho}+\Delta_0^{\nu\rho}\xi^{\mu}_{,\rho}\right)\right]
=-2\int\;d^4x'\;G^{\mu\nu\rho\sigma}_{,\sigma}\left[x,x'\right]\xi_{\rho}\left[x'\right]
\te
$\dot \xi^{\mu}=U^{\nu}\xi^{\mu}_{;\nu}$, so
\be
G^{\mu\nu\rho\sigma}_{,\sigma}\left[x,x'\right]=\frac12\rho_0\left[\eta^{\mu\rho}\left(U^{\nu}U^{\lambda}+\frac13\Delta_0^{\nu\lambda}\right)+\eta^{\nu\rho}\left(U^{\mu}U^{\lambda}+\frac13\Delta_0^{\mu\lambda}\right)\right]\partial_{\lambda}\delta\left(x-x'\right)
\te
Since the background is homogeneous, we expect the propagators to be translation invariant. Then we may Fourier transform
\be
G^{\mu\nu\rho\sigma}=\int\frac{d\omega d^3k}{\left(2\pi\right)^4}\;e^{i\left[\mathbf{k}\cdot\left(\mathbf{x-x'}\right)-\omega\left(t-t'\right)\right]}G^{\mu\nu,\rho\sigma}\left[\mathbf{k},\omega\right]
\te
whereby
\be
i\omega G^{\mu\nu\rho 0}\left[\mathbf{k},\omega\right]=ik_jG^{\mu\nu\rho j}\left[\mathbf{k},\omega\right]-\frac i2\rho_0\left[\eta^{\mu\rho}\left(U^{\nu}\omega-\frac13\Delta_0^{\nu j}k_j\right)+\eta^{\nu\rho}\left(U^{\mu}\omega-\frac13\Delta_0^{\mu j}k_j\right)\right]
\te
This means that $G^{\mu\nu\rho 0}$ is trivially obtained from $G^{\mu\nu,jk}$, and so there is no loss of generality in computing the propagators under the gauge condition $h_{\mu 0}=0$. 

On general grounds we expect the propagators will obey the reciprocity condition \cite{DGM,LLSP}
\be
G^{\mu\nu\rho\sigma}\left[x,x'\right]=G^{\rho\sigma\mu\nu}\left[x',x\right]
\label{reciprocity}
\te
whereby we also do not need to compute explicitly the propagators of the form $G^{0\nu\rho\sigma}$. So the only nontrivial problem is to compute the variation of $T^{ij}$ upon a metric fluctuation $h_{ij}$. However, to complete this task we need some information on the fundamental degrees of freedom of the theory and their dynamics.

In kinetic theory \cite{CalHu08,CanCal20,Isr72}, the fundamental description is provided by the distribution function $f\left(x^{\mu},p_{\mu}\right)$, where $p^{\mu}$, for a conformal theory, is restricted to the future light cone $p^2=0$, $p^0\ge 0$. $f$ obeys the Boltzmann equation
\be
p^{\mu}f_{;\mu}=I_{col}\left[f\right]
\label{Boltzmann}
\te
where the phase space covariant derivative is
\be
f_{;\mu}=\frac{\partial f}{\partial x^{\mu}}+\Gamma^{\nu}_{\mu\rho}p_{\nu}\frac{\partial f}{\partial p_{\rho}}
\te
Covariant derivatives are taken with the first order connection
\be
\Gamma^{\mu}_{\nu\lambda}=\frac12\eta^{\mu\rho}\left[h_{\nu\rho,\lambda}+h_{\lambda\rho,\nu}-h_{\nu\lambda,\rho}\right]
\te
Observe that $\Gamma^{\nu}_{\nu\lambda}=h^i_{i,\lambda}/2$ as expected,
and $\Gamma^{\mu}_{\nu\lambda}U^{\nu}U^{\lambda}=0$. 

For simplicity, we shall only consider the Maxwell-J\"uttner case where the equilibrium distribution has the form $f_{eq}=e^{\beta^{\mu}p_{\mu}}$, $\beta^{\mu}=u^{\mu}/T$ as above. The collision integral $I_{col}$ vanishes in equilibrium. It also has to satisfy the constraint
\be
\int\;Dp\;p^{\mu}I_{col}\left[f\right]=0
\te
with the phase space covariant measure
\be
Dp=\frac{4\pi d^4p_{\mu}}{\left(2\pi\right)^4\sqrt{-g}}\delta\left(-p^2\right)\theta\left(p^0\right)
\te
which enforces conservation eq. (\ref{cons}) for the EMT
\be
T^{\mu\nu}\left[f\right]=\int\;Dp\;p^{\mu}p^{\nu}\;f,
\label{kintmunu}
\te
and the $H$ theorem
\be
H=\int\;Dp\;\ln \left[f\right]\;I_{col}\left[f\right]\le 0
\label{Eta}
\te
for any solution of the Boltzmann equation (\ref{Boltzmann}). Validity of the $H$ theorem (\ref{Eta}) enforces the Second Law eq. (\ref{secondlaw}) with the entropy flux
\be
S^{\mu}\left[f\right]=\int\;Dp\;p^{\mu}\;f\left[1-\ln f\right]
\label{kins}
\te
and entropy production $\Sigma =-H$.  

To a given $f$ we may associate an EMT eq. (\ref{kintmunu}) and thereby a velocity $u^{\mu}$ and an energy density $\rho\equiv \sigma T^4$ through the Landau-Lifshitz prescription (\ref{LL}). We then adopt the relaxation time approximation \cite{AnWit74a,AnWit74b,BGK,TI10}

\be
I_{col}=\frac1{\tau}u^{\mu}p_{\mu}\left[f-f_{eq}\right]
\label{AW}
\te
where $f_{eq}$ is the Maxwell-J\"uttner distribution with the same $T$ and $u^{\mu}$ as $f$. The constant $\tau$ is the so called relaxation time.

In equilibrium $T=T_0$, $u^{\mu}=U^{\mu}=\left(1,0,0,0\right)$, $g^{\mu\nu}=\eta^{\mu\nu}$ and $f=f_0=f_{eq,0}=e^{-p^0/T_0}$. We consider a metric fluctuation $g_{\mu\nu}=\eta_{\mu\nu}+h_{\mu\nu}$ with $h_{0\nu}=0$. Subsequently we have $T=T_0\left(1+\vartheta\right)$ and $u^{\mu}=U^{\mu}+v^{\mu}$ with $v^0=0$. We parameterize
\be
f=f_0\left[1+\vartheta\frac {p^0}{T_0}+v^k\frac{p_k}{T_0}+\varphi\right]
\te
where $\vartheta$ and $v^{\mu}$ are the perturbations to the Landau-Lifshitz temperature and velocity, namely
\be
\int\;\frac{d^3p}{\left(2\pi\right)^3p}\left(U_{\nu}+v_{\nu}\right)p^{\nu}p^{\mu}f=-\sigma T_0^4\left[\left(1+4\vartheta\right)U^{\mu}+v^{\mu}\right]
\te
which is equivalent to
\be
\int\;\frac{d^3p}{\left(2\pi\right)^3}p^{\mu}\varphi f_0=0
\label{constraint}
\te
Then also
\be
f_{eq}=f_0\left[1+\vartheta\frac {p^0}{T_0}+v^k\frac{p_k}{T_0}\right]
\te
To perform the scalar-vector-tensor decomposition we write 
\begin{equation}
    v_i = ik_i v_S + v_{V i}
		\label{vdec}
\end{equation}
with $k_j v_{V j} = 0$, and 

\begin{equation}
    h_{ij} =  \left[   k_i k_j - k^2 \delta_{ij}\right] h_S + \left[ k_i k_j - \frac{1}{3} \delta_{ij} k^2 \right] h'_S +i k_i h_{V j} +i k_j h_{V i} + h_{T ij}
		\label{hdec}
\end{equation}
with $k_j h_{V j} = k_j h_{T ij} = h_{T jj} = 0$. $\vartheta$, $v_S$, $h_S$ and $h'_S$ correspond to scalar degrees of freedom, $v_V$ and $h_V$ are vector degrees of fredom, and $h^{jk}_T$ are the tensor degrees of freedom. The Boltzmann equation (\ref{Boltzmann}) becomes
\be
\frac p{T_0}\left(p\vartheta_{,t}+p^k\vartheta_{,k}\right)+\frac {p_j}{T_0}\left(pv^j_{,t}+p^kv^j_{,k}\right)+p\varphi_{,t}+p^k\varphi_{,k}+\frac1{2T_0}\dot h_{jk}p^{j}p^{k}=\frac{-p}{\tau}\varphi
\label{ktbe}
\te
Fourier transforming the space-time dependence, we get
\be
\varphi=\frac {p}{T_0}\frac{\left[-\frac {\omega}k+\hat k_l\hat p^l\right]\left(\vartheta +\hat p_jv_j\right)-\frac{\omega }{2k} h_{jk}\hat p^{j}\hat p^{k}}{z-\hat k_l\hat p^l}
\label{varphi}
\te
where
\be
z=\frac1k\left[\omega+i\tau^{-1}\right]
\label{zeta}
\te
On the other hand, if we multiply eq. (\ref{ktbe}) by $f_0$ and integrate over all momenta, the terms containing $\varphi$ cancel because of eq. (\ref{constraint}), and we get the continuity equation
\be
4\sigma T_0^4\left\{-i\omega\vartheta-\frac13k^2v_S+\frac i3\omega k^2h_S\right\}=0
\label{cont}
\te
We still must enforce the constraints eq. (\ref{constraint}), which become
\bea
\left[-\frac{\omega}kJ+\hat k_lJ^l\right]\vartheta+\left[-\frac{\omega}kJ^k+\hat k_lJ^{lk}\right]v_k&=&\frac{\omega}{2k} h_{jk}J^{jk}\nn
\left[-\frac{\omega}kJ^j+\hat k_lJ^{jl}\right]\vartheta+\left[-\frac{\omega}kJ^{jk}+\hat k_lJ^{jlk}\right]v_k&=&\frac{\omega}{2k} h_{lk}J^{jkl}
\label{kteom}
\tea
where
\be
J^{k_1\ldots k_n}=\int\;\frac{d^3p}{\left(2\pi\right)^3}\frac{p^2\hat p^{k_1}\ldots\hat p^{k_n}}{z-\hat k_l\hat p^l}\;e^{-p/T_0}
\label{integrals}
\te
They are evaluated in Appendix (\ref{integraleval}). After finding $\vartheta$, $v^k$ and $\varphi$, we may proceed to compute the EMT
\be
T^{\mu\nu}=\rho T_0^4\left(1+4\vartheta\right)\left[U^{\mu}U^{\nu}+\frac13\Delta^{\mu\nu}+U^{\mu}v^{\nu}+U^{\nu}v^{\mu}\right]+\Pi^{\mu\nu}
\te
where
\be
\Pi^{\mu\nu}=\int\;\frac{d^3p}{\left(2\pi\right)^3p}\;p^{\nu}p^{\mu}\varphi\;f
\te
or else
\be
\Pi^{jk}=\frac 1{T_0}\left\{\left[-\frac{\omega}kJ^{jk}+\hat k_lJ^{jkl}\right] \vartheta +\left[-\frac{\omega}kJ^{jkm}+\hat k_lJ^{jklm}\right]v_m-\frac{\omega}{2k} h_{lm}J^{jklm}\right\}
\label{pijk}
\te
Observe that the trace of $\Pi^{jk}$ vanishes because the $J^{k_1\ldots k_n}$ tensors obey $J^{j k_1\ldots k_n}_j=J^{k_1\ldots k_n}$.

It is clear that the scalar, vector and tensor sectors decouple, and it is best to consider each one in turn. We expect to recover the Landau-Lifshitz hydrodinamic modes from divergences in $\vartheta$ and $v^j$, while there will be poles in $\varphi$ associated to non hydrodynamic modes absent from Landau-Lifshitz theory.

\subsection{Hydrodynamic poles}

\subsubsection{The scalar sector}

In the scalar sector we have
\begin{equation}
    h_{ij} =  \left[   k_i k_j - k^2 \delta_{ij}\right] h_S + \left[ k_i k_j - \frac{1}{3} \delta_{ij} k^2 \right] h'_S 
\end{equation}
and $v^k=iv_Sk^k$. From the results of appendix (\ref{integraleval}) we get
\bea
\left[-\frac{\omega}kJ+\chi_1\right]\vartheta+\left[-\frac{\omega}k\chi_1+\frac23\chi_3\right]ikv_S&=&\frac{\omega k}{3}\left[\left(-\frac32J+\chi_3\right)h_S+\left(-\frac12J+\chi_3\right)h'_S\right] \nn
\left[-\frac{\omega}k\chi_1+\frac23\chi_3\right]\vartheta+\left[-\frac23\frac{\omega}k\chi_3+\frac25\chi_5\right]ikv_S&=&\frac{\omega k}{5} \left[\left(2\chi_5-5\chi_1\right)h_S+\left(2\chi_5-\frac53\chi_1\right)h'_S\right] 
\label{SC}
\tea
We expect to find the hydrodynamic poles when the determinant of this system vanishes. We consider the $\tau\to 0$ limit, where $\left|z\right|\gg 1$. The condition for a vanishing determinant, to next to lowest order in $z^{-1}$, is
\be
\omega\approx\pm\frac k{\sqrt{3}}-\frac 2{15}i\tau k^2
\te
which reproduces the result for Landau-Lifshitz fluids eq. (\ref{LLdisp}),  identifying $\nu=\tau/5$.

\subsubsection{The vector sector}

In the vector sector
\begin{equation}
    h_{ij} =  i k_i h_{V j} +i k_j h_{V i} 
\end{equation}
with $h_{Vi}k^i=0$. Then $\vartheta=0$ and $v^i=v_V^i$ is also transverse. We find
\be
\left[-\frac{\omega}k\left(\frac12J-\frac13\chi_3\right)+\frac12\chi_1-\frac15\chi_5\right]v_V^k=\frac{i\omega}{2} \left(\chi_1-\frac25\chi_5\right)h_V^k
\label{VC}
\te
There is a pole when the term in brackets in the left hand side vanishes. For large $z$ (cfr. eq. (\ref{zeta})) we find the pole at
\be
\omega\approx\frac1{5}\frac{-ik^2\tau}{\left(1-\omega\tau\right)}
\label{KTVHP}
\te 
which corresponds to the hydrodynamic pole we find in Landau-Lifshitz fluids, again identifying $\nu=\tau/5$ (cfr. eq. (\ref{LLvec})).

In the tensor sector $h_{ij}=h_{Tij}$ is both traceless and divergenceless, $\vartheta=v^i=0$, and there are no hydrodynamic poles.

\subsection{Non hydrodynamic poles}

To study the non hydrodynamic modes, we shall consider $\Pi^{jk}$ in the limit where $k\to 0$, while $\omega$ goes to a finite value and $\left|z\right|\gg 1$. 
Once again, it is best to consider scalar, vector and tensor modes separatedly.

\subsubsection{Scalar modes}

In the limit we are considering, the scalar viscous EMT reads
\be
\Pi_S^{jk}=\frac {-\omega}{T_0k}\left\{J^{jk} \vartheta+J^{jkm}\hat k_m\left(ikv_S\right)+\frac12k^2\left[\left(h_S+h'_S\right)J^{jklm}\hat k_l\hat k_m-\left(h_S+\frac13h'_S\right)J^{jklm}\delta_{lm}\right]\right\}
\label{pisjk}
\te
Since $\Pi^j_{Sj}=0$ we may write
\be
\Pi_S^{jk}=\Pi_S\left[\hat k^j\hat k^k-\frac13\delta^{jk}\right]
\te
\bea
\Pi_S&=&\frac32\Pi_S^{jk}\hat k_j\hat k_k\nn
&=&\frac {-\omega}{T_0k}\left\{\chi_3\vartheta+\frac35\chi_5\left(ikv_S\right)+\frac12k^2\left[\left(h_S+h'_S\right)\left(-\frac32J+2\chi_3+12\chi_6\right)-\left(h_S+\frac13h'_S\right)\chi_3\right]\right\}
\label{pis}
\tea
It may be seen that to this order there are no new poles proportional to $h_S$, so we shall set $h_S=0$. Then
we only need the leading order form of $ikv_S$ and $\vartheta$. In the scalar case eq. (\ref{cont}) and the two equations (\ref{kteom}) are not independent, we choose to work with the former and the first of the latter. Then
\be
ikv_S=-\frac{ 2k^2}{15z} h'_S 
\te
while $\vartheta$ is of higher order in $k$. From eq. (\ref{pis}) we now find
\be
\Pi_S=-\frac {\sigma T_0^4\omega k}{z}\frac4{15}\left(1+\frac9{35}\frac1{z^2}\right)h'_S
\te
We see that the leading order terms in the inverse propagator go like $1-\left(9/35z^2\right)$. This behavior may be reproduced by a non hydrodynamic pole at $z^2=9/35$, or else
\be
\omega=\frac{-i}{\tau}\pm\sqrt{\frac 9{35}}k
\label{ktnhs}
\te

\subsubsection{Vector modes}

As with the scalar modes, we begin with the leading order relation

\be
v_V^k\approx\frac{-i k}{5z}h_V^k
\label{VCnh}
\te
The vector viscous EMT reduces to
\be
\Pi_V^{jk}=\frac {-{4i\omega}}{15zT_0}\left\{1+\frac8{35}\frac{1}{z^2} \right\}\left(\hat k^j h_{V}^k+\hat k^k h_{V}^j\right)
\label{pijknh}
\te
Therefore the leading terms in the inverse propagator are $\propto 1-(8/35)z^{-2}$, ``as if'' there were a non hydrodynamic pole at 
\be
\omega=-\frac i{\tau}\pm\sqrt{\frac8{35}}k
\label{ktnhv}
\te

\subsubsection{Tensor modes}

The response of the EMT to a tensor metric fluctuation is
\be
\Pi_T^{jk}=\frac {-\omega}{ T_0k}\chi_6h_{T}^{jk}
\te
For large $z$ we get
\be
\Pi_T^{jk}=\frac {-4i\omega\tau}{ 15\left(1-i\omega\tau\right)}\sigma T_0^4h_{T}^{jk}\left[1+\frac17\frac1{z^2}+\ldots\right]
\te
Therefore the inverse propagator will be proportional to $1-1/7z^2$, which is the behavior caused by a pole at \cite{Roma666}
\be
\omega=-\frac i{\tau}\pm\frac k{\sqrt{7}}
\label{ktnht}
\te
We see from kinetic theory that a relativistic fluid may support damped tensor waves, which are totally absent in first order theories. 

\subsection{Poles or cuts?}\label{AdSCFT}

The analytic structure of the kinetic theory propagators is determined by the presence of a cut in the complex frecuency plane. This is due to the fact that the propagators depend on the function $L\left[z\right]$ defined in eq. (\ref{elezeta}), which has a cut from $z=-1$ to $z=1$ or else, with $z$ as in eq. (\ref{zeta}), from $\omega=-i/\tau-k$ to $-i/\tau+k$ \cite{KurWie}. However, the analytic structure of the propagators in the $k\to 0$ limit may be reproduced by suitably locating poles in the complex $\omega$ plane. Though the poles themselves are not in the asymptotic region, they give the right analytic structure there.

Coincidentally, the analytic structure of EMT correlators in $\lambda\phi^4$ theory is determined by cuts, rather than poles \cite{Moore}.

However, in the strong coupling limit things seem to be different. We can verify this explicitly in theories with a holographic dual \cite{Rom-Rom-19,BRSSS,PolSonStar,SonStarBlackMagic,PolSonStar2,PolSonStar3,HerSon,KotSonStar,KotStar,ChesYaf,Withers,Grozdanov}. Then the thermal EMT correlators may be found by solving the classical Einstein equations in a dual, five dimensional space time. A thermal state corresponds to an Anti de Sitter (AdS) space time containing a black hole in its center. The physical fluid lives on the boundary of the AdS space time, and the Hawking temperature of the hole is the temperature of the fluid. Perturbations of the fluid correspond to perturbations of the black hole metric, with no naked singularities and incoming boundary conditions at the horizon.

For each $k$, the frequencies of free oscillations of the fluid correspond to the horizon fluctuations, and they come in a discrete, infinite series, depending on the t'Hooft coupling $g^2N_c$, where $g$ and $N_c$ are the coupling and number of colors of the underlying gauge theory \cite{Hooft2}. These free frequencies appear as poles in the EMT propagators. For the scalar (``sound'') and vector (``shear'') sectors, there is sequence of hydrodynamic poles, namely $\omega\to 0$ when $k\to 0$ and also nonhydrodynamic poles. In the infinite coupling limit, the hydrodynamic poles are well reproduced by a FOT with $\nu=4\pi/T_0$. Corrections to this value at finite coupling are discussed in \cite{Waeber15}. In the tensor (in this context frequently called ``scalar'') sector, there are no hydrodynamic poles, but again an infinite, discrete sequence of non hydrodynamic ones. 

The presence of infinitely many modes cannot be reproduced by hydrodynamics, either first or second order. However, the imaginary part of the poles increases rather sharply along the sequence, and so only a few long lived modes are actually relevant to describe the approach to equilibrium. These are the modes that either causal FOTs or second order theories aim to reproduce.

\subsection{Beyond the relaxation time approximation}

While the relaxation time approximation we have used so far is frequently a preferred choice in view of its simplicity, it is also known \cite{Noronha21} that it departs in significant ways from more realistic kinetic equations such as Boltzmann's and Landau's. It is therefore relevant to ask how far results derived under this approximation may be generalized to more complete setups, if at all.

To this end we adopt the viewpoint presented in \cite{JPR2010}. We parameterize
\be
f=f_0\left(1+\phi\right)
\te
And 
\be
I_{col}\left[f\right]=-\left(p^0\right)f_0\mathcal{F}\left[\phi\right]
\te
The Boltzmann equation (\ref{Boltzmann}) becomes
\be
\phi_{,t}+\hat p^k\phi_{,k}+\frac1{2T_0p^0}\dot h_{jk}p^{j}p^{k}=-\mathcal{F}\left[\phi\right]
\label{ktbe}
\te
We assume the collision term is ultralocal in position space. Then it commutes with Fourier transformation, and we get
\be
\left[\omega-k_k\hat p^k\right]\phi+\frac{\omega}{2T_0p^0}h_{jk}p^{j}p^{k}=-i\mathcal{F}\left[\phi\right]
\label{ktbeft}
\te
It is convenient to introduce an inner product in the space of functions of momentum
\be
\left\langle \psi\vert\phi\right\rangle=\int\;D_{\beta}p\;\psi^*\phi
\te
where
\be
D_{\beta}p=Dp\;p^0f_0
\te
We assume the linearized operator $\mathcal{F}$ is symmetric under this inner product. Momentum conservation takes the form
\be
\left\langle p^{\mu}\left|\mathcal{F}\right|\phi\right\rangle=0
\te
Since $\left|\phi\right\rangle$ can be any vector, it must be
\be
\mathcal{F}\left|p^{\mu}\right\rangle=0
\te
We assume these are the only null eigenvectors of the collision operator, and that there is a set of non-null eigenvectors
\be
\mathcal{F}\left|\phi_{\lambda}\right\rangle=\nu_{\lambda}\left|\phi_{\lambda}\right\rangle
\te
where $\nu_{\lambda}$ is real and positive, and $\left\langle \phi_{\sigma}\vert\phi_{\lambda}\right\rangle=\delta_{\sigma\lambda}$. We then have
\bea
\left|\phi\right\rangle&=&\sum_{\mu=0}^3\alpha_{\mu}\left|p^{\mu}\right\rangle\left\langle p^{\mu}\vert\phi\right\rangle+\sum_{\lambda}\left|\phi_{\lambda}\right\rangle\left\langle \phi_{\lambda}\vert\phi\right\rangle\nn
\mathcal{F}\left|\phi\right\rangle&=&\sum_{\lambda}\nu_{\lambda}\left|\phi_{\lambda}\right\rangle\left\langle \phi_{\lambda}\vert\phi\right\rangle
\tea
where (no sum over $\mu$)
\be
\alpha_{\mu}=\left\langle p^{\mu}\vert p^{\mu}\right\rangle^{-1}
\te
Contracting eq. (\ref{ktbeft}) with the $\left|p^{\mu}\right\rangle$ or the $\left|\phi_{\lambda}\right\rangle$ we get two sets of equations. On one hand

\be
\sum_{\nu=0}^3\left[\omega\delta^{\mu\nu}-k_k\alpha_{\nu}\left\langle p^{\mu}\left|\hat p^k\right|p^{\nu}\right\rangle\right]\left\langle p^{\nu}\vert\phi\right\rangle-\sum_{\sigma}k_k\left\langle p^{\mu}\left|\hat p^k\right|\phi_{\sigma}\right\rangle\left\langle \phi_{\sigma}\vert\phi\right\rangle=-\frac{\omega}{2T_0}h_{jk}\left\langle p^{\mu}\vert\frac{p^{j}p^{k}}{p^0}\right\rangle
\te
and on the other
\be
-\sum_{\nu=0}^3k_k\alpha_{\nu}\left\langle \phi_{\lambda}\left|\hat p^k\right|p^{\nu}\right\rangle\left\langle p^{\nu}\vert\phi\right\rangle+\sum_{\sigma}\left[\left(\omega +i\nu_{\sigma}\right)\delta_{\lambda\sigma}-k_k\left\langle \phi_{\lambda}\left|\hat p^k\right|\phi_{\sigma}\right\rangle\right]\left\langle \phi_{\sigma}\vert\phi\right\rangle=-\frac{\omega}{2T_0}h_{jk}\left\langle \phi_{\lambda}\vert\frac{p^{j}p^{k}}{p^0}\right\rangle
\te
where

\bea
\alpha_0&=&\frac13\alpha_i=2\pi^2\left[T_0^5\Gamma\left[5\right]\right]^{-1}\nn
\left\langle p^{0}\left|\hat p^k\right|p^{0}\right\rangle&=&\left\langle p^{j}\left|\hat p^k\right|p^{l}\right\rangle=\left\langle p^{l}\vert\frac{p^{j}p^{k}}{p^0}\right\rangle=0\nn
\left\langle p^{j}\left|\hat p^k\right|p^{0}\right\rangle&=&\left\langle p^{0}\left|\hat p^k\right|p^{j}\right\rangle=\left\langle p^{0}\vert\frac{p^{j}p^{k}}{p^0}\right\rangle=\frac{T_0^5\Gamma\left[5\right]}{6\pi^2}\delta^{jk}\nn
\tea
We then get

\bea
\left\langle p^{0}\vert\phi\right\rangle&=&\frac{k_k}{\omega}\left\langle p^{k}\vert\phi\right\rangle+\frac{k^2h_{S}}{3T_0\alpha_0}\nn
\left\langle p^{i}\vert\phi\right\rangle&=&\frac{k^i}{3\omega}\left\langle p^{0}\vert\phi\right\rangle+\sum_{\sigma}\frac{k_k}{\omega}\left\langle \frac{p^{i}p^k}{p^0}\vert\phi_{\sigma}\right\rangle\left\langle \phi_{\sigma}\vert\phi\right\rangle\nn
\left\langle \phi_{\lambda}\vert\phi\right\rangle&=&-\frac{\omega h_{jk}}{2T_0\left(\omega +i\nu_{\lambda}\right)}\left\langle \phi_{\lambda}\vert\frac{p^{j}p^{k}}{p^0}\right\rangle+\frac{k_k}{\left(\omega +i\nu_{\lambda}\right)}\left\{\sum_{l}3\alpha_{0}\left\langle \phi_{\lambda}\vert\frac{p^kp^{l}}{p^0}\right\rangle\left\langle p^{l}\vert\phi\right\rangle+\sum_{\sigma}\left\langle \phi_{\lambda}\left|\hat p^k\right|\phi_{\sigma}\right\rangle\left\langle \phi_{\sigma}\vert\phi\right\rangle\right\}\nn
\tea

Since we are only interested in the hydrodynamic limit, we may solve these equations in powers of $k^j$. To zeroth order

\bea
\left\langle p^{0}\vert\phi\right\rangle^{\left(0\right)}&=&k^2h_{S}\frac{T_0^4\Gamma\left[5\right]}{6\pi^2}\nn
\left\langle p^{j}\vert\phi\right\rangle^{\left(0\right)}&=&0\nn
\left\langle \phi_{\lambda}\vert\phi\right\rangle^{\left(0\right)}&=&-\frac{\omega h_{jk}}{2T_0\left[\omega +i\nu_{\lambda}\right]}\left\langle \phi_{\lambda}\vert\frac{p^{j}p^{k}}{p^0}\right\rangle
\tea

$\left\langle p^{0}\vert\phi\right\rangle$ receives no first order corrections. Else

\bea
\left\langle p^{j}\vert\phi\right\rangle^{\left(1\right)}&=&\frac{1}{\omega}\left\{k^{j} k^2h_{S}\frac{T_0^4\Gamma\left[5\right]}{18\pi^2}-\sum_{\sigma}k_k\left\langle \frac{p^{j}p^k}{p^0}\vert\phi_{\sigma}\right\rangle\frac{\omega h_{lm}}{2T_0\left[\omega +i\nu_{\sigma}\right]}\left\langle \phi_{\sigma}\vert\frac{p^{l}p^{m}}{p^0}\right\rangle\right\}\nn
\left\langle \phi_{\lambda}\vert\phi\right\rangle^{\left(1\right)}&=&-\frac{k_k}{\left(\omega +i\nu_{\lambda}\right)}\sum_{\sigma}\left\langle \phi_{\lambda}\left|\hat p^k\right|\phi_{\sigma}\right\rangle\frac{\omega h_{lm}}{2T_0\left[\omega +i\nu_{\sigma}\right]}\left\langle \phi_{\sigma}\vert\frac{p^{l}p^{m}}{p^0}\right\rangle
\tea

To second order we get

\bea
\left\langle p^{0}\vert\phi\right\rangle^{\left(2\right)}&=&\frac{k_k}{\omega}\left\langle p^{k}\vert\phi\right\rangle^{\left(1\right)}\nn
\left\langle p^{j}\vert\phi\right\rangle^{\left(2\right)}&=&\frac{k_k}{\omega}\sum_{\sigma}\left\langle \frac{p^{j}p^k}{p^0}\vert\phi_{\sigma}\right\rangle\left\langle \phi_{\sigma}\vert\phi\right\rangle^{\left(1\right)}\nn
\left\langle \phi_{\lambda}\vert\phi\right\rangle^{\left(2\right)}&=&\frac{k_k}{\left(\omega +i\nu_{\lambda}\right)}\left\{\sum_{l}\alpha_{l}\left\langle \phi_{\lambda}\vert\frac{p^kp^{l}}{p^0}\right\rangle\left\langle p^{l}\vert\phi\right\rangle^{\left(1\right)}+\sum_{\sigma}\left\langle \phi_{\lambda}\left|\hat p^k\right|\phi_{\sigma}\right\rangle\left\langle \phi_{\sigma}\vert\phi\right\rangle^{\left(1\right)}\right\}
\tea

Finally, the spatial components of the perturbed EMT

\bea
T^{ij}-T_0^{ij}&=&\left\langle \frac{p^ip^j}{p^0}\vert\phi\right\rangle\nn
&=&\frac13\delta^{ij}\left\{k^2h_{S}\frac{T_0^4\Gamma\left[5\right]}{6\pi^2}\left(1+\frac{k^2}{3\omega^2}\right)-\frac{k_kk_l}{\omega^2}\sum_{\sigma}\left\langle \frac{p^{k}p^l}{p^0}\vert\phi_{\sigma}\right\rangle\frac{\omega h_{lm}}{2T_0\left[\omega +i\nu_{\sigma}\right]}\left\langle \phi_{\sigma}\vert\frac{p^{l}p^{m}}{p^0}\right\rangle\right\}\nn
&+&\sum_{\lambda}\left\langle \frac{p^ip^j}{p^0}\vert\phi_{\lambda}\right\rangle\left\{-\frac{\omega h_{lm}}{2T_0\left[\omega +i\nu_{\lambda}\right]}\left\langle \phi_{\lambda}\vert\frac{p^{l}p^{m}}{p^0}\right\rangle\right.\nn
&+&\left.\frac{k_k}{\left(\omega +i\nu_{\lambda}\right)}\left\{\sum_{l}3\alpha_{0}\left\langle \phi_{\lambda}\vert\frac{p^kp^{l}}{p^0}\right\rangle\frac{1}{\omega}\left\{k^{l} k^2h_{S}\frac{T_0^4\Gamma\left[5\right]}{18\pi^2}-\sum_{\sigma}k_m\left\langle \frac{p^{l}p^{m}}{p^0}\vert\phi_{\sigma}\right\rangle\frac{\omega h_{pq}}{2T_0\left[\omega +i\nu_{\sigma}\right]}\left\langle \phi_{\sigma}\vert\frac{p^{p}p^{q}}{p^0}\right\rangle\right\}\right.\right.\nn
&-&\left.\left.\sum_{\sigma}\left\langle \phi_{\lambda}\left|\hat p^k\right|\phi_{\sigma}\right\rangle\left\{\frac{k_l}{\left(\omega +i\nu_{\sigma}\right)}\sum_{\sigma'}\left\langle \phi_{\sigma}\left|\hat p^l\right|\phi_{\sigma'}\right\rangle\frac{\omega h_{pq}}{2T_0\left[\omega +i\nu_{\sigma'}\right]}\left\langle \phi_{\sigma'}\vert\frac{p^{p}p^{q}}{p^0}\right\rangle\right\}\right\}\right\}
\label{fullmonty}
\tea

We observe that there must be eigenvectors for which $\left\langle \left(p^ip^j/p^0\right)\vert\phi_{\lambda}\right\rangle\not= 0$, since $\left|\left(p^ip^j/p^0\right)\right\rangle$ is not a linear superposition of the null eigenvectors $\left|p^{\mu}\right\rangle$ and the $\left|\phi_{\lambda}\right\rangle$ are complete in the complement of the $\left|p^{\mu}\right\rangle$. In coordinates where $k^i$ is in the $z$-direction, this is true in particular for $\left|p^xp^y/p^0\right\rangle$, which couples to the tensor part of $h_{ij}$, $\left|p^xp^z/p^0\right\rangle$, which couples to the vector part, and $\left|\left(p^{x2}+p^{y2}-2p^{z2}\right)/p^0\right\rangle$, which couples to the scalar part. So there must be a nontrivial propagator for tensor modes just as there is for vector and scalar ones. Also, if the collision term does not break the isotropy of space, we expect the eigenvectors to carry a representation of the rotation group, and for this reason at least, the eigenvectors that couple to these kets (and therefore cannot be rotation invariant) will be degenerate.

Once we have the representation (\ref{fullmonty}) at our disposal, it is a simple matter to match it to the asympotic behavior resulting from a suitable distribution of poles and cuts in the complex frequency plane. For example, in the first term of eq. (\ref{fullmonty}) we find

\be
\left(1+\frac{k^2}{3\omega^2}\right)\approx\frac{\omega^2}{\omega^2-\left(k^2/3\right)}
\te
signaling the presence of the usual sound pole. Then it is easy to see that eigenvalues in the discrete spectrum will give rise to poles, while eigenvalues in the continuous spectrum will be associated to cuts. This derives from the fact that an expression such as 

\be
\int_{\nu_0}^{\infty}d\nu\;\frac{\rho\left[\nu\right]}{\left[\omega +i\nu\right]}
\te
is discontinuous when we go from $\omega=-i\nu+\epsilon$ to $\omega=-i\nu-\epsilon$, for all $\nu>\nu_0$ such that $\rho\left[\nu\right]\not=0$.

The relaxation time approximation is the extreme case where the full spectrum is collapsed to just two points, $0$ and $1/\tau$. The sums over the non-null eigenvalues are performed by using that

\be
\sum_{\sigma}\left\langle \psi'\vert\phi_{\sigma}\right\rangle\left\langle \phi_{\sigma}\vert\psi\right\rangle=\left\langle \psi'\vert\psi\right\rangle-\sum_{\nu=0}^3\alpha_{\nu}\left\langle \psi'\vert p^{\nu}\right\rangle\left\langle  p^{\nu}\vert\psi\right\rangle
\te
and we easily recover the results above. In the opposite limit, the theories with an energy dependent relaxation time discussed  in \cite{KurWie} (see also \cite{Noronha21,JPR2010,Luzum,Dusling}) have purely continuous spectrum ranging from $\nu_0=0$ to $\infty$, and therefore lead to an analytic structure dominated by cuts.

Realistic kinetic equations such as Boltzmann's \cite{DeGroot,Stewart,Cercignani}  and Landau's \cite{LLPK,Martinez2020} have both isolated and continuous eigenvalues, strongly dependent upon the details of the interparticle interactions. The so-called ``hard'' potentials have a continuos spectrum ranging from a finite value $\nu_0>0$ to $\infty$ \cite{Liboff}; they may have further isolated eigenvalues between the everpresent $0$ and $\nu_0$ \cite{Dudynski,LuoYu}, and besides, since most modes in the continuum decay much faster than $\nu_0^{-1}$, it may be possible, for all practical purposes, to approximate the continuum by a single eigenvalue. Contrarywise, ``soft'' interactions lead to a continuous spectrum starting from $\nu_0=0$ \cite{Liboff}. We should note that both Boltzmann's and Landau's equations have a fifth null eigenvector associated with particle number conservation, which is not a feature of conformal theories (in gauge theories there are particle number changing processes such as gluon splitting, in $\lambda\phi^4$ theory particle number conservation is broken at order $\lambda^4$ \cite{CHR2000}). More general collision terms are discussed in \cite{CalHu08}.

We therefore conclude that the analytic structure found under the relaxation time approximation is not an artifact of the approximation, although it is not universal either, and its validity must be judged on a case by case basis.

%%%%%%%%%%%%%%%%%%%%%%%%%%%%%%%%%%%%%%%%%%%%%%%%%%%%%%%%%%%%%%
%%%%%%%%%%%%%%%%%%%%%%%%%%%%%%%%%%%%%%%%%%%%%%%%%%%%%%%%%%%%%%
%%%%%%%%%%%%%%%%%%%%%%%%%%%%%%%%%%%%%%%%%%%%%%%%%%%%%%%%%%%%%%

\section{Dispersion relations in Divergence type theories}\label{DTT}

In this section we shall compute the dispersion relations as derived from SOTs within a DTT scheme.

Let us motivate this particular SOT by deriving it from kinetic theory \cite{CanCal20}. To this end we assume a parameterized distribution function
\be
f=f_{hydro}\left[x^{\mu},p_{\mu};\beta^{\mu},\zeta^{\mu\nu},\ldots\right]
\te
We obtain equations of motion for the parameters $\beta^{\mu},\zeta^{\mu\nu},\ldots$ by taking suitable averages of the Boltzmann equation \cite{DMNR12,DNMR12}
\bea
\int\;Dp\;p^{\mu}p^{\nu}\;f_{hydro;\nu}&=&\int\;Dp\;p^{\mu}I_{col}\left[f_{hydro}\right]=0\nn
\int\;Dp\;R_{\alpha}\left[x,p\right]\;p^{\nu}f_{hydro;\nu}&=&\int\;Dp\;R_{\alpha}I_{col}\left[f_{hydro}\right]
\ldots
\label{averages}
\tea
This scheme enforces energy-momentum conservation eq. (\ref{cons}). However, because $f_{hydro}$ is not a solution of the Boltzmann equation (\ref{Boltzmann}), the $H$ theorem ec. (\ref{Eta}) does not apply to it, and the Second Law may not be enforced. This is avoided if we make the specific parameterization
\be
f_{hydro}=e^{\beta^{\mu}p_{\mu}+\sum_{\alpha}\zeta^{\alpha}R_{\alpha}\left[x,p\right]}
\te
that is, the same functions $R_{\alpha}$ which appear in $f_{hydro}$ are averaged against the Boltzmann equation to obtain the equations of motion for the parameters $\zeta^{\alpha}$  \cite{CanCal20}. The resulting system of equations take the form of conservation laws for the currents
\be
A^{\mu}_{\alpha}=\int\;Dp\;R_{\alpha}\left[x,p\right]\;p^{\mu}f_{hydro}
\te
and so this theory falls within the class of DTTs.

The particular implementation of this scheme we are interested in postulates two nonequilibrium tensors $\zeta_{\mu\nu}$ and $\xi_{\mu\nu\rho}$ besides the usual variables $T$ and $u^{\mu}$. The parameterized distribution function reads
\be
f_{hydro}=e^{\beta_{\mu}p^{\mu}+\zeta_{\mu\nu}\frac{p^{\mu}p^{\nu}}{\left(-u_{\lambda}p^{\lambda}\right)}+\xi_{\mu\nu\rho}\frac{p^{\mu}p^{\nu}p^{\rho}}{\left(-u_{\lambda}p^{\lambda}\right)^2}}
\te
where $\beta^{\mu}=u^{\mu}/T$, $u^2=-1$. $\zeta_{\mu\nu}$ and $\xi_{\mu\nu\rho}$ are totally symmetric, traceless and transverse, meaning that
\bea
S^{\mu\nu}_{\sigma\lambda}\zeta^{\sigma\lambda}&=&\zeta^{\mu\nu}\nn
S^{\mu\nu\rho}_{\sigma\lambda\tau}\xi^{\sigma\lambda\tau}&=&\xi^{\mu\nu\rho}
\tea
with the projectors
\bea
S^{\mu\nu}_{\sigma\lambda}&=&\frac12\left\{\Delta^{\mu}_{\sigma}\Delta^{\nu}_{\lambda}+\Delta^{\mu}_{\lambda}\Delta^{\nu}_{\sigma}-\frac23\Delta^{\mu\nu}\Delta_{\sigma\lambda}\right\}\nn
S^{\mu\nu\rho}_{\sigma\lambda\tau}&=&\frac16\left\{\Delta^{\mu}_{\sigma}\Delta^{\nu}_{\lambda}\Delta^{\rho}_{\tau}+\Delta^{\mu}_{\sigma}\Delta^{\nu}_{\tau}\Delta^{\rho}_{\lambda}+\Delta^{\mu}_{\lambda}\Delta^{\nu}_{\sigma}\Delta^{\rho}_{\tau}\right.\nn
&+&\Delta^{\mu}_{\lambda}\Delta^{\nu}_{\tau}\Delta^{\rho}_{\sigma}+\Delta^{\mu}_{\tau}\Delta^{\nu}_{\lambda}\Delta^{\rho}_{\sigma}+\Delta^{\mu}_{\tau}\Delta^{\nu}_{\sigma}\Delta^{\rho}_{\lambda}\nn
&-&\frac25\left[\Delta^{\mu\nu}\left(\Delta^{\rho}_{\sigma}\Delta_{\lambda\tau}+\Delta^{\rho}_{\lambda}\Delta_{\sigma\tau}+\Delta^{\rho}_{\tau}\Delta_{\lambda\sigma}\right)\right.\nn
&+&\Delta^{\mu\rho}\left(\Delta^{\nu}_{\sigma}\Delta_{\lambda\tau}+\Delta^{\nu}_{\lambda}\Delta_{\sigma\tau}+\Delta^{\nu}_{\tau}\Delta_{\lambda\sigma}\right)\nn
&+&\left.\left.\Delta^{\nu\rho}\left(\Delta^{\mu}_{\sigma}\Delta_{\lambda\tau}+\Delta^{\mu}_{\lambda}\Delta_{\sigma\tau}+\Delta^{\mu}_{\tau}\Delta_{\lambda\sigma}\right)\right]\right\}
\tea
where as usual $\Delta^{\mu\nu}=g^{\mu\nu}+u^{\mu}u^{\nu}$. 

We should note that most work on DTTs to date does not include the third order tensor $\xi_{\mu\nu\rho}$. Often the only variables considered are a chemical potential (for non conformal theories), the four vector $\beta_{\mu}$ and a traceless tensor $\zeta_{\mu\nu}$ \cite{statisticalDTT}. This adds up to fourteen degrees of freedom, and is thus analog to Grad's ``fourteen moments'' approximation. In these usual theories the tensor mode in non propagating, while, as we shall show below, including the $\xi_{\mu\nu\rho}$ tensor  provides it with a finite propagation speed, which further agrees with the one derived from kinetic theory eq. (\ref{ktnht}).

Including the tensor $\xi_{\mu\nu\rho}$ will not just produce this only change in the theory, we should expect there will be incremental changes in the scalar and vector sectors as well. However it is fair to say that they do not change the physical picture in those sectors as they do for the tensor modes. Likewise, including higher order tensors will only have incremental effects on $T^{\mu\nu}$.

The equations of motion for the variables $\beta^{\mu}$, $\zeta^{\mu\nu}$ and $\xi^{\mu\nu\rho}$ are obtained from the weighted averages of the covariant Boltzmann equation (\ref{Boltzmann})
\bea
\int\;Dp\;p^{\mu}\left[p^{\tau}f_{hydro;\tau}-I_{col}\right]&=&0\nn
\int\;Dp\;S^{\alpha\beta}_{\mu\nu}\frac{p^{\mu}p^{\nu}}{\left(-u_{\lambda}p^{\lambda}\right)}\left[p^{\tau}f_{hydro;\tau}-I_{col}\right]&=&0\nn
\int\;Dp\;S^{\alpha\beta\gamma}_{\mu\nu\rho}\frac{p^{\mu}p^{\nu}p^{\rho}}{\left(-u_{\lambda}p^{\lambda}\right)^2}\left[p^{\tau}f_{hydro;\tau}-I_{col}\right]&=&0
\tea
Integrating by parts, we write these equations as conservation laws
\bea
T^{\mu\nu}_{;\nu}&=&0\nn
S^{\alpha\beta}_{\mu\nu}\left[A^{\mu\nu\rho}_{;\rho}-A^{\mu\nu\rho\sigma}u_{\rho;\sigma}-I^{\mu\nu}\right]&=&0\nn
S^{\alpha\beta\gamma}_{\mu\nu\rho}\left[A^{\mu\nu\rho\sigma}_{;\sigma}-2A^{\mu\nu\rho\sigma\lambda}u_{\sigma;\lambda}-I^{\mu\nu\rho}\right]&=&0
\tea
where
\bea
T^{\mu\nu}&=&\int\;Dp\;p^{\mu}p^{\nu}\;f_{hydro}\nn
A^{\mu_1\ldots\mu_n}&=&\int\;Dp\;\frac{p^{\mu_1}\ldots p^{\mu_n}}{\left(-u_{\lambda}p^{\lambda}\right)^{n-2}}\;f_{hydro}\nn
I^{\mu_1\ldots\mu_n}&=&\int\;Dp\;\frac{p^{\mu_1}\ldots p^{\mu_n}}{\left(-u_{\lambda}p^{\lambda}\right)^{n-1}}\;I_{col}
\tea
If $I_{col}$ satisfies the $H$ theorem, then this dynamics yields positive entropy production, with the entropy flux
\bea
S^{\mu}&=&\int\;Dp\;p^{\mu}f_{hydro}\left[1-\ln f_{hydro}\right]\nn
&=&\Phi^{\mu}-\beta_{\nu}T^{\mu\nu}-\zeta_{\nu\rho}A^{\mu\nu\rho}-\xi_{\nu\rho\sigma}A^{\mu\nu\rho\sigma}
\tea
and entropy production
\be
\Sigma=-\zeta_{\nu\rho}I^{\nu\rho}-\xi_{\nu\rho\sigma}I^{\nu\rho\sigma}
\te
Here
\be
\Phi^{\mu}=\int\;Dp\;p^{\mu}\;f_{hydro}
\te
So far the discussion has been general. We now linearize around an equilibrium solution.
We adopt the relaxation time approximation collision integral eq. (\ref{AW}).

In equilibrium $\zeta^{\mu\nu}=\xi^{\mu\nu\rho}=0$; since we are interested in linear deviations from equilibrium only, we can write
\be
f_{hydro}=e^{\beta_{\mu}p^{\nu}}\left[1+\zeta_{\mu\nu}\frac{p^{\mu}p^{\nu}}{\left(-u_{\lambda}p^{\lambda}\right)}+\xi_{\mu\nu\rho}\frac{p^{\mu}p^{\nu}p^{\rho}}{\left(-u_{\lambda}p^{\lambda}\right)^2}\right]
\te
then also $f_{eq}=e^{\beta_{\mu}p^{\nu}}$ and the equations of motion reduce to
\bea
T^{\mu\nu}_{0;\nu}+T_{1}^{\mu\nu\rho\sigma}\zeta_{\rho\sigma;\nu}&=&0\nn
S^{\alpha\beta}_{\mu\nu}\left[T^{\mu\nu\rho}_{1;\rho}+T_{1}^{\mu\nu\rho\sigma}\left(\dot\zeta_{\rho\sigma}+\frac1{\tau}\zeta_{\rho\sigma}\right)+T^{\mu\nu\rho\sigma\lambda\tau}_{3}\xi_{\sigma\lambda\tau;\rho}-T_{2}^{\mu\nu\rho\sigma}u_{\rho;\sigma}\right]&=&0\nn
S^{\alpha\beta\gamma}_{\mu\nu\rho}T^{\mu\nu\rho\sigma\lambda\tau}_{3}\left[\dot\xi_{\sigma\lambda\tau}+\frac1{\tau}\xi_{\sigma\lambda\tau}+\zeta_{\lambda\sigma;\tau}\right]&=&0
\tea
where
\be
T_{\alpha}^{\mu_1\ldots\mu_n}=\int\;Dp\;\frac{p^{\mu_1}\ldots p^{\mu_n}}{\left(-u_{\lambda}p^{\lambda}\right)^{\alpha}}\;f_{eq}
\label{tensors}
\te
These tensors are evaluated in Appendix (\ref{tensoreval}). In general, the projectors mean that we have to symmetrize and subtract all longitudinal and tracefull terms. We also write $T=T_0\left(1+\vartheta\right)$ and $u^{\mu}=U^{\mu}+v^{\mu}$, with $U^{\mu}v_{\mu}=0$. We then get
\bea
\dot\vartheta+\frac13\left(v^i_{,i}+\frac12\dot h^i_i\right)&=&0\nn
\dot v_j+\vartheta_{,j}+\frac 2{5}T_0\zeta^k_{j,k}&=&0\nn
\frac12\sigma_{jk}+T_0\left(\dot\zeta_{jk}+\frac1{\tau}\zeta_{jk}\right)+\frac37T_0\xi_{jkl,l}&=&0\nn
\frac13\left[\zeta_{ij,k}+\zeta_{ik,j}+\zeta_{jk,i}-\frac25\left(\delta_{ij}\zeta_{kl,l}+\delta_{ik}\zeta_{jl,l}+\delta_{kj}\zeta_{il,l}\right)\right]+\dot\xi_{ijk}+\frac1{\tau}\xi_{ijk}&=&0
\label{theseequations}
\tea
where $\sigma_{ij}$ is the shear tensor

\be
\sigma_{jk}=v_{i,j}+v_{j,i}-\frac23\delta_{ij}v^{k}_{,k}+\dot h_{ij}-\frac13\delta_{ij}\dot h^k_k
\label{shear}
\te
Fourier transforming and using the velocity decomposition (\ref{vdec}) and the metric decomposition (\ref{hdec}) we get
\bea
\sigma_{jk}&=&-2\left[k_ik_j-\frac13\delta_{ij}k^2\right]v_S+i\left(k_iv_{Vj}+k_jv_{Vi}\right)\nn
&-&i\omega\left[\left[ k_i k_j - \frac{1}{3} \delta_{ij} k^2 \right] \left(h_S+h'_S\right)+i k_i h_{V j} +i k_j h_{V i} + h_{T ij}\right]
\label{shearsvt}
\tea

To decompose eqs. (\ref{theseequations}) in scalar, vector and tensor modes, we define
\bea
\zeta_{ij}&=&\zeta_S\left[k_ik_j-\frac13k^2\delta_{ij}\right]+i\left(\zeta_{Vi}k_j+\zeta_{Vj}k_i\right)+\zeta_{Tij}\nn
\xi_{ijk}&=&i\xi_S\left[k_ik_jk_k-\frac15k^2\left(k_i\delta_{jk}+k_j\delta_{ik}+k_k\delta_{ij}\right)\right]\nn
&+&\xi_{Vi}\left[k_jk_k-k^2\frac15\delta_{jk}\right]+\xi_{Vj}\left[k_ik_k-k^2\frac15\delta_{ik}\right]+\xi_{Vk}\left[k_jk_i-k^2\frac15\delta_{ji}\right]\nn
&+&i\left(\xi_{Tij}k_k+\xi_{Tik}k_j+\xi_{Tkj}k_i\right)+\xi_{TTijk}
\tea
where $k_i\zeta_{Vi}=k_i\zeta_{Tij}=0$, $k_i\xi_{Vi}=k_i\xi_{Tij}=k_i\xi_{TTijk}=0$, and tensors are totally symmetric and traceless. The scalar-vector-tensor decomposition of the shear tensor is given in eq. (\ref{shearsvt}).

\subsection{The scalar sector}

In the scalar sector we get
\bea
i\omega\vartheta+\frac13k^2v_S&=&\frac i3\omega k^2 h_S\nn
\vartheta-i\omega v_S+\frac{4}{15}k^2T_0\zeta_S&=&0\nn
v_S+i\left(\omega+\frac i{\tau}\right)T_0\zeta_S+\frac9{35}k^2T_0\xi_S&=&\frac12i\omega\left(h_S+h'_S\right)\nn
\zeta_S-i\left(\omega+\frac i{\tau}\right)\xi_S&=&0
\tea
The dispersion relation is
\be
\left(\omega^2-\frac13k^2\right)\left(\omega+\frac i{\tau}\right)^2-k^2\left[\frac{4}{15}\omega\left(\omega+\frac i{\tau}\right)+\frac9{35}\left(\omega^2-\frac 13k^2\right)\right]=0
\te
When $k^2,\tau\to 0$ we have two branches of solutions,  hydrodynamic modes with
\be
\omega\approx\pm\frac k{\sqrt{3}}-\frac 2{15}i\tau k^2
\te
which correspond to the Landau-Lifshitz modes with the identification $\nu=\tau/5$ (see Appendix (\ref{CFOTs})),
and nonhydrodynamic modes with
\be
\omega\approx\frac{-i}{\tau}\pm \sqrt{\frac9{35}} k
\te
just as derived from kinetic theory, eq. (\ref{ktnhs}).

\subsection{The vector sector}

In the vector sector we get
\bea
\omega v_{Vj}-\frac2{5}ik^2T_0\zeta_{Vj}&=&0\nn
\frac12iv_{Vj}+\left(\omega+\frac i{\tau}\right)T_0\zeta_{Vj}+\frac{12}{35}ik^2T_0\xi_{Vj}&=&-\frac12i\omega h_{Vj}\nn
\frac23i\zeta_{Vj}-\left(\omega+\frac i{\tau}\right)\xi_{Vj}&=&0
\tea
The dispersion relation is 
\be
\omega\left(\omega+\frac i{\tau}\right)^2-\frac1{5}k^2\left(\omega+\frac i{\tau}\right)-\frac{8}{35}k^2\omega=0
\te
Therefore when $k^2\to 0$, either $\omega\to 0$ or $\omega\to -i\tau^{-1}$. In the first case we find an hydrodynamic mode with
\be
\omega\approx -\frac1{5}i\tau k^2
\te
while the others are two nonhydrodynamic modes with
\be
\omega\approx\frac{-i}{\tau}\pm\sqrt{\frac{8}{35}}k
\te
Once again, the hydrodynamic modes agree with Landau-Lifshitz theory if $\nu=\tau/5$ (see Appendix (\ref{CFOTs})), and also reproduces the nonhydrodynamic mode from kinetic theory eq. (\ref{ktnhv}). 

\subsection{The tensor sector}

In the tensor sector we get
\bea
\left(\omega+\frac i{\tau}\right)\zeta_{Tjk}-\frac37ik^2\xi_{Tjk}&=&\frac{-i}{T_0}\omega h_{Tjk}\nn
\frac13i\zeta_{Tjk}+\left(\omega+\frac i{\tau}\right)\xi_{Tjk}&=&0
\tea
We therefore find two nontrivial hydrodynamic modes with 
\be
\omega\approx\frac{-i}{\tau}\pm\sqrt{\frac{1}{7}}k
\te
These modes have no analog in Landau-Lifshitz fluids, but match quantitatively the nonhydrodinamic tensor modes from kinetic theory eq. (\ref{ktnht}).

If we had considered a truncated theory with $\xi_{ijk}=0$ then there would be a tensor nonhydrodynamic mode, but with a $k-$ independent dispersion relation $\omega=-i/\tau$, thus not propagating. They are present already in Israel-Stewart theory \cite{HisLind3,Natsuume08}. Their cosmological consequences are discussed in \cite{Nahuel18,Nahuel21}.

$\xi_{TTijk}$ is decoupled and obeys the equation
\be
\left(\omega+\frac i{\tau}\right)\xi_{TTijk}=0
\te
so we may assume it vanishes throughout.

\section{Final remarks}
In this paper we have computed the dispersion relations for a DTT containing two non equilibrium tensors $\zeta_{\mu\nu}$ and $\xi_{\mu\nu\rho}$ besides the usual variables $T$ and $u^{\mu}$, and compared the result to known results in the literature involving FOTs \cite{Nor1} and also ``first principles'' calculations from kinetic theory \cite{Roma666,KurWie,Heller18,Heller21} and quantum field theory \cite{Rom-Rom-19,Moore,BRSSS,PolSonStar,SonStarBlackMagic,PolSonStar2,PolSonStar3,HerSon,KotSonStar,KotStar,ChesYaf,Withers,Grozdanov}.

The ``first principles'' calculations display both hydrodynamic and nonhydrodynamic modes in all three sectors, scalar vector and tensor. Both FOTs and DTTs describe well the hydrodynamic modes and the longest living non hydrodynamic modes in the scalar and vector sectors. DTTs also describe the longest living tensor mode, which is not recovered in FOTs. A truncated DTT with only the $\zeta_{\mu\nu}$ tensor yields a non propagating tensor mode, in agreement with Israel-Stewart theory \cite{HisLind3,Natsuume08}.

We have been unable to find a clear cut statement about the speed of propagation of tensor modes in the third order formalism \cite{Greiner2010,BHMR08,Grozdanov2016,thirdorder}, since usually only the ``sound'' and ``shear'' channels (which correspond to scalar and vector modes in this paper) are discussed in detail. The relevant third order equation as derived in \cite{J13} and \cite{J15} includes second order derivatives of the viscous EMT.

DTTs also perform well in the similar problem of the dispersion relations for a viscous charged fluid \cite{Biswas2020,Biswas2021}, where they reproduce the Weibel instability \cite{weibel59,CalKan16}. The cosmological consequences of this fact are discussed in \cite{KMGC21}. 

We believe these results validate the choice of DTTs to describe fluids in problems where the interaction with gravitational waves is a matter of relevance \cite{Maggiore}. These problems include generation and amplification of gravitational waves in the Very Early Universe \cite{Caprini,Nahuel18,Nahuel21} and during cosmological phase transitions \cite{Hindmarsh}, and gravitational wave emission from rotating neutron stars \cite{Friedman}, among others. Where gravitational interactions are not a concern, the relative advantages of FOTs and SOTs must be considered carefully to find the most suitable model for each application.

\section*{Acknowledgments}
We thank M. Nigro, A. Kandus, N. Mir\'on Granese, L. Cantarutti and Juli\'an Ruffinelli for multiple discussions.
This work  was supported in part by Universidad de Buenos Aires through
grant UBACYT 20020170100129BA, CONICET and ANPCyT.

\appendix

\section{Dispersion relations for ideal and Landau-Lifshitz fluids, and in causal FOTS}\label{CFOTs}

\subsection{Ideal fluids}

To obtain the dispersion relations for an ideal fluid we must consider the conservation laws eq. (\ref{cons}) for the EMT eq. (\ref{ideal}). For a conformal fluid, $\rho=\sigma T^4$ and $p=\rho/3$. For concreteness we assume the value of the constant $\sigma$ which is compatible with Maxwell-J\"uttner statistics for a single degree of freedom, namely $\sigma=3/\pi^2$. The conservation laws are
\bea
\frac{\dot T}T+\frac13u^{\nu}_{;\nu}&=&0\nn
\dot u^{\mu}+\Delta^{\mu\nu}\frac{T_{,\nu}}T&=&0
\tea
$\dot X=u^{\mu}X_{;\mu}$. We linearize these equations writing $T=T_0\left(1+\vartheta\right)$, $u^{\mu}=U^{\mu}+v^{\mu}$, $U^{\mu}=\left(1,0,0,0\right)$, and $v^0=0$. 
Taking the Fourier transform we get
\bea
-i\omega\vartheta+\frac13\left[ik_jv^j-\frac i2\omega h^i_i\right]&=&0\nn
-i\omega v^j+ik_j\vartheta&=&0
\tea
We now decompose the velocity as in eq. (\ref{vdec} and the metric as in eq. (\ref{hdec}).
No tensor degrees of freedom are included.
 We thereby get on one hand
\bea
-\omega\vartheta+\frac i3k^2v_S&=&-\frac1{3}\omega k^2h_S\nn
\vartheta-i\omega v_S&=&0
\tea
and on the other
\be
\omega v^j_V=0
\te
We see that scalar, vector and tensor modes decouple. For the scalar modes we find the dispersion relation
\be
\omega^2-\frac13k^2=0
\te
so the scalar propagator has poles at $\omega=\pm k/\sqrt{3}$, representing longitudinal sound waves.  No vector or tensor modes are excited in the fluid.

\subsection{Landau-Lifshitz fluids}

If we consider a Landau-Lifshitz fluid instead of an ideal one, the difference is that the EMT becomes
\be
T^{\mu\nu}=\frac43\sigma T^4\left[u^{\mu}u^{\nu}+\frac14g^{\mu\nu}\right]-\eta\sigma^{\mu\nu}
\label{deltatjk}
\te
where $\eta$ is the fluid viscosity and $\sigma^{\mu\nu}$ is the shear tensor
\be
\sigma^{\mu\nu}=\Delta^{\mu\lambda}\Delta^{\nu\rho}\left[u_{\lambda;\rho}+u_{\rho;\lambda}-\frac23\Delta_{\lambda\rho}u^{\sigma}_{;\sigma}\right]
\label{covshear}
\te
The linearized conservation equations are now
\bea
\dot \vartheta+\frac13v^{j}_{;j}&=&0\nn
\dot v^{j}+\vartheta_{,j}-\nu\sigma^{jk}_{;k}&=&0
\tea
where $\nu=3\eta/4\sigma T_0^4=\eta/ s_0T_0$ is the kinematic viscosity. With the shear tensor eq. (\ref{shear}) and its Fourier transform eq. (\ref{shearsvt}) we get
\bea
-\omega\vartheta+\frac i3k^2v_S&=&-\frac1{3}\omega k^2h_S\nn
\vartheta-i\left[\omega+\frac{4i}3\nu k^2\right] v_S&=&\frac12i k^2\nu\omega\left(h_S+h'_S\right)
\tea
and 
\be
\left[\omega+i\nu k^2\right] v^j_V=-\frac34\nu k^2\omega h^j_V
\label{LLvec}
\te
We  see that the longitudinal sound waves now become damped, with a dispersion relation
\be
\omega\approx\pm\frac k{\sqrt{3}}-\frac23 i\nu k^2
\label{LLdisp}
\te
while the pole for transverse waves has been shifted from $\omega=0$ to $\omega=-i\nu k^2$. There are still no tensor modes in the fluid.

\subsection{Causal FOTs}

As a representative causal FOT we shall consider the EMT \cite{Nor1}
\begin{equation}
\begin{split}
    T^{\mu \nu} = &\left[ \epsilon + \frac{3 \chi}{4 \epsilon} \left( u^{\rho} \nabla_{\rho} \epsilon + \frac{4}{3} \epsilon \nabla_{\rho} u^{\rho} \right) \right] \left( u^{\mu} u^{\nu} + \frac13{\Delta^{\mu \nu}} \right)-\eta\sigma^{\mu\nu} \\
    &+ \frac{\lambda}{4 \epsilon} \left[4 u^{\mu} \epsilon u^{\rho} \nabla_{\rho} u^{\nu} + u^{\mu} \Delta^{\nu}_{\rho} \nabla^{\rho} \epsilon + 4 u^{\nu} \epsilon u^{\rho} \nabla_{\rho} u^{\mu} + u^{\nu} \Delta^{\mu}_{\rho} \nabla^{\rho} \epsilon \right]
\end{split}
\end{equation}
{Only when in equilibrium, $\epsilon=\rho$  is the energy density. The new transport coeffcients $\chi$ and $\lambda$ are the hallmark of the causal FOT approach. They define time scales of the order of the relaxation time on which the hydrodynamic variable $\epsilon$ relaxes to the Landau-Lifshitz energy density, and in this sense they act as causal regulators. See \cite{Nor1} for further discussion.}  

To linearize the conservation equations (\ref{cons}) we write $\epsilon = \epsilon_0 + \delta \epsilon$, and expand $u^{\mu}=U^{\mu}+v^{\mu}$ and $g^{\mu \nu} = \eta^{\mu \nu} + h^{\mu \nu}$ as before. We further decompose $v^{i}$ as in eq. (\ref{vdec}) and $h_{ij}$ as in eq. (\ref{hdec}). We thus get the Fourier-transformed equations
\begin{equation}
    0 = - i\omega \left[ \delta \epsilon + \frac{3 \chi}{4 \epsilon_0} \left( -i\omega \delta \epsilon - \frac{4}{3} \epsilon_0 k^2 v_S \right) \right] - \left[ \frac{4}{3} \epsilon_0 - \lambda i\omega \right] k^2 v_S - \frac{1}{6} \epsilon_0 i\omega h_S
\end{equation}

\bea
\label{Eq:Space}
    0 &=& -i\omega \left[ \frac{4}{3} \epsilon_0 - \lambda i\omega \right] \left( ik^i v_S + v_V^i \right) + \frac{1}{3} \epsilon_0 \left( -i\frac{2}{3} k^2 k^i h'_S - k^2 h_V^i \right) + \frac{1}{6} i\epsilon_0 k^i k^2 h_S \nn
    &+& \frac{1}{3} ik^i \left[ \delta \epsilon + \frac{3 \chi}{4 \epsilon_0} \left( -i\omega \delta \epsilon - \frac{4}{3} \epsilon_0 k^2 v_S \right) \right] - \frac{1}{3} \epsilon_0 \left[ \frac{2}{3} ik^2 k^i h'_S + k^2 h_V^i \right] \nn
    &+& 2 \eta \left[ \frac{1}{2} \left( k^2 (ik^i v_S + v_V^i) + ik^i ( k^2 v_S ) - i\omega k^2 ik^i h'_S \right) - \frac{1}{3} ik^i \left(- k^2 v_S + i\omega k^2 h_S \right) \right]
\tea
We already see that the tensor sector has trivial dynamics. We analyze the scalar and vector sectors in turn.

\subsubsection{Scalar sector}

The scalar part of eq. (\ref{Eq:Space}) reads

\bea
    0 &=& -i\omega \left[ \frac{4}{3} \epsilon_0 - \lambda i\omega \right] v_S - \frac{4}{9} \epsilon_0 k^2 h'_S + \frac{1}{6} \epsilon_0 k^2 h_S + \frac{1}{3} \left[ \delta \epsilon + \frac{3 \chi}{4 \epsilon_0} \left( -i\omega \delta \epsilon - \frac{4}{3} \epsilon_0 k^2 v_S \right) \right] \nn
    &+& 2 \eta k^2 \left[ \frac{2}{3} v_S - \frac{1}{2} i\omega h'_S + \frac{1}{3} i\omega h_S \right]
\tea
Elliminating  $v_S$ we find
\bea
    \frac{\delta \epsilon}{\epsilon_0} && \left[ 1 - \frac{\epsilon_0k^2}{3i\omega}   \frac{\left[ \frac{4}{3} - i\omega \left( \frac{\lambda + \chi}{\epsilon_0} \right) \right]}{\left[ -\omega^2 \lambda - i\omega \frac{4}{3} \epsilon_0 - \frac{1}{3} k^2 \left( \chi - 4 \eta \right) \right]}\right] = - \frac{1}{6} \frac{1}{\left[1 - i\omega \frac{3 \chi}{4 \epsilon_0}\right]} h_S \nn
    &+&\frac{ k^2}{i\omega} \frac{\left[ \frac{4}{3} - i\omega \left( \frac{\lambda + \chi}{\epsilon_0} \right) \right]}{1 -i \omega \frac{3 \chi}{4 \epsilon_0}} \left\{ \frac{\frac{2}{3} \epsilon_0 k^2  \left( \frac{2}{3} h'_S - \frac{1}{4} h_S \right) - \eta i\omega k^2 \left( \frac{2}{3} h_S - h'_S \right)}{-\omega^2 \lambda - i\omega \frac{4}{3} \epsilon_0 + \frac{1}{3} k^2 \left( \chi - 4 \eta \right)} \right\}
\label{BDN}
\tea
In the limit $k^2\to 0$ eq. (\ref{BDN}) simplifies to
\be
\frac{\delta \epsilon}{\epsilon_0}=-\frac1{\left[1 - i\omega \frac{3 \chi}{4 \epsilon_0}\right]}\frac1{\left[\omega^2-\frac13k^2\left(1-\alpha\right)\right]}\frac{h_S}6
\te
where
\be
\alpha=\frac34i\omega\frac{\chi}{\epsilon_0}\left[1-\frac34i\omega\frac{\lambda}{\epsilon_0}\right]^{-1}
\te
We therefore get a hydrodynamic pole with
\be
\omega=\pm\frac k{\sqrt{3}}-\frac i8\frac{\chi}{\epsilon_0}k^2
\te
and a non hydrodynamic pole with
\be
\omega=-i\frac{4 \epsilon_0}{3 \chi}
\te

\subsubsection{Vector sector} 

From the vector part of eq. (\ref{Eq:Space}) we get
\begin{equation}
\begin{split}
    v_V^i = \frac{2}{3}  \left( \frac{-\epsilon_0 k^2}{-\eta k^2+ \frac{4}{3} i\omega \epsilon_0 + \lambda \omega^2 } \right) h^i_V
\end{split}
\end{equation}
When $k^2\to 0$ we get a hydrodynamic mode
\be
\omega=-\frac34i\frac{\eta}{\epsilon_0}k^2
\te
and a non hydrodynamic mode
\be
\omega=-\frac43i\frac{\epsilon_0}{\lambda}+\frac34i\frac{\eta}{\epsilon_0}k^2
\te
We see that the causal FOT formalism recovers the hydrodynamic scalar and vector modes, and also non hydrodynamic modes in these sectors, with no dynamics at all in the tensor sector.

\section{Evaluation of integrals}\label{integraleval}
The integrals eq. (\ref{integrals}) may be evaluated explicitly by going to polar coordinates with $\hat k_l\hat p^l=\cos\theta=x$
\be
J=\frac1{4\pi^2}\int_0^{\infty}dp\;p^4e^{-p/T_0}\int_{-1}^1\frac{dx}{z-x}=2\sigma T_0^5L\left[z\right]
\te
where
\be
L\left[z\right]=\ln\frac{z+1}{z-1}=2\sum_{n=0}^{\infty}\frac1{\left(2n+1\right)}\frac1{z^{2n+1}}
\label{elezeta}
\te
The radial integral is always the same, while the $x$ integral changes. We have
\be
J^k=\chi_1\hat k^k
\te
\bea
\chi_1&=&2\sigma T_0^5\int_{-1}^1\frac{dx\;x}{z-x}\nn
&=&2\sigma T_0^5\left[zL\left[z\right]-2\right]=\frac43\sigma T_0^5\left[\frac1{z^2}+\frac35\;\frac1{z^4}+\ldots\right]
\tea
To evaluate $J^{jk}$ write
\be
J^{jk}=\chi_2\left[\hat k^j\hat k^k-\delta^{jk}\right]+\chi_3\left[\hat k^j\hat k^k-\frac13\delta^{jk}\right]
\te
Then
\be
\chi_2=-\frac12J^j_j=-\frac12J=-2\sigma T_0^5\left[\frac1z+\frac13\frac1{z^3}+\ldots\right]
\te
\bea
\chi_3&=&3\sigma T_0^5\int_{-1}^1\frac{dx\;x^2}{z-x}\nn
&=&6\sigma T_0^5\left[\frac12z^2L\left[z\right]-z\right]=2\sigma T_0^5\left[\frac1z+\frac35\frac1{z^3}+\ldots\right]
\tea
A totally symmetric third order tensor may be written as
\be
J^{ijk}=a\hat k^i\hat k^j\hat k^k+b\left(\hat k^i\delta^{jk}+\hat k^j\delta^{ki}+\hat k^k\delta^{ij}\right)
\te
A traceless tensor must have $a+5b=0$, while a totally divergenceless tensor must have $a+3b=0$. So we may also write
\be
J^{ijk}=\chi_4\left[\hat k^i\hat k^j\hat k^k-\frac13\left(\hat k^i\delta^{jk}+\hat k^j\delta^{ki}+\hat k^k\delta^{ij}\right)\right]+\chi_5\left[\hat k^i\hat k^j\hat k^k-\frac15\left(\hat k^i\delta^{jk}+\hat k^j\delta^{ki}+\hat k^k\delta^{ij}\right)\right]
\te
Now $J^{ij}_j=J^i$, so
\be
\chi_4=-\frac32\chi_1=-2\sigma T_0^5\left[\frac1{z^2}+\frac35\;\frac1{z^4}+\ldots\right]
\te
and then
\bea
\chi_5&=&5\sigma T_0^5\int_{-1}^1\frac{dx\;x^3}{z-x}\nn
&=&5\sigma T_0^5\left[z^3L\left[z\right]-2z^2-\frac23\right]=2\sigma T_0^5\left[\frac1{z^2}+\frac57\;\frac1{z^4}+\ldots\right]
\tea
Finally we compute $J^{ijkl}$. Start from
\bea
J^{ijkl}&=&a\;\hat k^i\hat k^j\hat k^k\hat k^l\nn
&+&b\left(\hat k^i\hat k^j\delta^{kl}+\hat k^i\hat k^k\delta^{lj}+\hat k^i\hat k^l\delta^{jk}+\hat k^j\hat k^k\delta^{il}+\hat k^j\hat k^l\delta^{ik}+\hat k^k\hat k^l\delta^{ij}\right)\nn
&+&c\left(\delta^{ij}\delta^{kl}+\delta^{ik}\delta^{lj}+\delta^{il}\delta^{jk}\right)
\tea
We then have
\be
J^{ijk}_k=\left(a+7b\right)\hat k^i\hat k^j+\left(b+5c\right)\delta^{ij}=J^{ij}
\te
Then
\bea
a+7b&=&\chi_2+\chi_3\nn
b+5c&=&-\chi_2-\frac13\chi_3
\tea
so
\bea
J^{ijkl}&=&\chi_2\left[8\hat k^i\hat k^j\hat k^k\hat k^l-\left(\hat k^i\hat k^j\delta^{kl}+\hat k^i\hat k^k\delta^{lj}+\hat k^i\hat k^l\delta^{jk}+\hat k^j\hat k^k\delta^{il}+\hat k^j\hat k^l\delta^{ik}+\hat k^k\hat k^l\delta^{ij}\right)\right]\nn
&+&\frac13\chi_3\left[10\hat k^i\hat k^j\hat k^k\hat k^l-\left(\hat k^i\hat k^j\delta^{kl}+\hat k^i\hat k^k\delta^{lj}+\hat k^i\hat k^l\delta^{jk}+\hat k^j\hat k^k\delta^{il}+\hat k^j\hat k^l\delta^{ik}+\hat k^k\hat k^l\delta^{ij}\right)\right]\nn
&+&\chi_6\left[\delta^{ij}\delta^{kl}+\delta^{ik}\delta^{lj}+\delta^{il}\delta^{jk}-5\left(\hat k^i\hat k^j\delta^{kl}+\hat k^i\hat k^k\delta^{lj}+\hat k^i\hat k^l\delta^{jk}+\hat k^j\hat k^k\delta^{il}+\hat k^j\hat k^l\delta^{ik}+\hat k^k\hat k^l\delta^{ij}\right)+35\hat k^i\hat k^j\hat k^k\hat k^l\right]\nn
\tea
and
\bea
2\chi_2+\frac43\chi_3+8\chi_6&=&2\sigma T_0^5\int_{-1}^1\frac{dx\;x^4}{z-x}\nn
&=&2\sigma T_0^5\left[z^4L\left[z\right]-2z^3-\frac23z\right]
\tea
therefore
\be
\chi_6=\frac12\sigma T_0^5\left[\frac12\left(z^2-1\right)^2L\left[z\right]-z^3+\frac{5}3z\right]=\frac4{15}\sigma T_0^5\left[\frac1z+\frac17\frac1{z^3}+\ldots\right]
\te

\section{The $T_{\alpha}^{\mu_1\ldots\mu_n}$ tensors}\label{tensoreval}

In this appendix we evaluate the $T_{\alpha}^{\mu_1\ldots\mu_n}$ tensors from eq. (\ref{tensors}). The integrals against $f_{eq}$ may be computed on symmetry, tracelessness and dimensional grounds

\bea
T_0^{\mu\nu}&=&\sigma T_0^4\left[u^{\mu}u^{\nu}+\frac13\Delta^{\mu\nu}\right]\nn
T_1^{\mu\nu\rho}&=&\sigma T_0^4\left[u^{\mu}u^{\nu}u^{\rho}+\frac13\left(\Delta^{\mu\nu}u^{\rho}+\ldots\right)\right]\nn
T_1^{\mu\nu\rho\sigma}&=&4\sigma T_0^5\left[u^{\mu}u^{\nu}u^{\rho}u^{\sigma}+\frac13\left(\Delta^{\mu\nu}u^{\rho}u^{\sigma}+\ldots\right)+\frac1{15}\left(\Delta^{\mu\nu}\Delta^{\rho\sigma}+\ldots\right)\right]\nn
T_2^{\mu\nu\rho\sigma}&=&\sigma T_0^4\left[u^{\mu}u^{\nu}u^{\rho}u^{\sigma}+\frac13\left(\Delta^{\mu\nu}u^{\rho}u^{\sigma}+\ldots\right)+\frac1{15}\left(\Delta^{\mu\nu}\Delta^{\rho\sigma}+\ldots\right)\right]\nn
T_3^{\mu\nu\rho\sigma\lambda\tau}&=&4\sigma T_0^5\left[u^{\mu}u^{\nu}u^{\rho}u^{\sigma}u^{\lambda}u^{\tau}+\frac13\left(\Delta^{\mu\nu}u^{\rho}u^{\sigma}u^{\lambda}u^{\tau}+\ldots\right)+\frac1{15}\left(\Delta^{\mu\nu}\Delta^{\rho\sigma}u^{\lambda}u^{\tau}+\ldots\right)\right.\nn
&+&\left.\frac1{105}\left(\Delta^{\mu\nu}\Delta^{\rho\sigma}\Delta^{\lambda\tau}+\ldots\right)\right]\nn
\tea

\end{document}